\documentclass[12pt]{iopart}

\usepackage{graphicx}
\usepackage{amssymb}
\usepackage{setstack}
\usepackage{times}
\usepackage{mathrsfs}

\begin{document}

\title{Stochastic multi-channel lock-in detection}
\author{J J Hudson, M R Tarbutt, B E Sauer, E A Hinds}

\address{Centre for Cold Matter, Physics Department, Imperial College London, London SW7~2AZ, UK}
\ead{jony.hudson@imperial.ac.uk}

\begin{abstract}
High-precision measurements benefit from lock-in detection of small signals. Here we discuss the extension of lock-in detection to many channels, using mutually orthogonal modulation waveforms, and show how the the choice of waveforms affects the information content of the signal. We also consider how well the detection scheme rejects noise, both random and correlated. We address the particular difficulty of rejecting a background drift that makes a reproducible offset in the output signal and we show how a systematic error can be avoided by changing the  waveforms between runs and averaging over many runs. These advances made possible a recent measurement of the electron's electric dipole moment\,\cite{Hudson11}.
\end{abstract}

\maketitle

\section{Introduction}
Lock-in detection is a well-known method for picking out a small modulated signal of interest in the presence of a large background.  After multiplying the signal-plus-background by the modulation waveform, the dc component is proportional to the signal while the background now alternates and can be filtered out. An early account of this idea dates back to 1934 \cite{Cosens34}, when it was used to detect the balanced state of an ac bridge. Subsequent improvements to the electronics by Michels, Curtis and Redding \cite{Michels41,Michels49} resulted in a practical, narrow-band detector of high sensitivity, now widely known as a lock-in amplifier \footnote{Michels coined the name ``lock-in amplifier'' in Ref.\,\cite{Michels41}, then tried to change it in Ref.\,\cite{Michels49} to ``synchronous detector'' but the original name is the one that has stuck.}. This became a ubiquitous laboratory tool after it was commercialised by  Robert Dicke and Princeton Applied Research \cite{Dicke85}.

The lock-in technique has been extended to measure several modulated signals simultaneously. In the context of high-precision measurement, Harrison, Player, and Sandars (HPS) \cite{Harrison71}  describe signal modulation with complex, multi-frequency waveforms chosen to be mutually orthogonal. More recently in the context of optical tomography refs.\ \cite{Schmitz02, Masciotti08, Trinker13} consider phase-sensitive detection at several frequencies. Careful choice of the frequencies and filter time constants allows the signals to be extracted independently.

In this paper we extend the work of HPS. First, in order to establish a theoretical framework, we review the case when a single parameter is modulated. Then, in section\,\ref{Sec:multi} we revist and extend the HPS discussion of multi-channel modulation and detection. In section\,\ref{Sec:repeat}, we introduce and analyse the problem of reproducible backgrounds and in section \ref{Sec:random} we present a new, stochastic technique that solves the problem. We close with a summary in section \ref{Sec:conclusions}.

\section{Single channel demodulation}
\label{Sec:single}

Consider a simple experiment with a scalar output signal, $s$, and a binary control parameter, $c$, which we set to the value $+1$ or $-1$ over each of $N$ equal time intervals, labeled by the integer $j$. We sometimes refer to the set of values $c_j$ as a switching pattern or waveform. We write the signal measured in each time interval as
\begin{equation}
s_j = \beta_j+ \alpha c_j\ ,
\end{equation}
with $j=0, \ldots N-1$.  Here $\beta_j$ is a background term, which may be time-varying, and $\alpha$ is the modulated part of the signal that we are trying to extract, which we will assume to be constant. Note that this form of the signal does not assume a linear response to the control parameter $c_j$. We obtain an estimate, $\hat\alpha$, of the modulation strength, $\alpha$, by looking at the part of the signal correlated with $c_j$,
\begin{equation}\label{eq:extract}
\hat\alpha = \frac{1}{N} \sum_{j = 0}^{N - 1}{s_j c_j}\ .
\end{equation}
The error in this estimate of $\alpha$ is,
\begin{equation}
\label{eq:demoderr}
r_\alpha = \hat\alpha - \alpha = \frac{1}{N} \sum_{j = 0}^{N - 1}{\beta_jc_j}\ . 
\end{equation}
Our goal is to devise a scheme that gives the best estimate of $\alpha$, least affected by the presence of the background $\beta$.
 
If $\beta$ is constant during an experiment, then any switching waveform that is equally often positive and negative will reject it perfectly. We will call a waveform that has this property \textit{balanced}. It is more realistic though to let the background vary in time, and then the cancellation of $\beta$ is normally imperfect. To optimise the choice of $c_j$ we must consider in more detail how $\beta_j$ varies.
Let us define the discrete Fourier transform of $\beta$,
\begin{equation}
(\mathscr{F}\beta)_k= \frac{1}{\sqrt{N}}\sum_{l=0}^{N-1}\beta_l e^{-i  \frac{2 \pi k}{N}l} \,,
\end{equation}
and write Eq.\,(\ref{eq:demoderr}) in terms of these Fourier components,
\begin{equation}
r_\alpha = \frac{1}{N} \sum_{j = 0}^{N - 1}{ \left( \frac{1}{\sqrt{N}}\sum_{k=0}^{N-1}(\mathscr{F}\beta)_k e^{i \frac{2 \pi k}{N}j} \right)  c_j}\ ,
\end{equation}
where the large-bracketed quantity is nothing more than $\beta_j$. Rewriting this as
\begin{equation}
\label{eq:fourier}
r_\alpha = \frac{1}{N} \sum_{k= 0}^{N - 1}{ (\mathscr{F}\beta)_k}  \left( \frac{1}{\sqrt{N}}\sum_{j=0}^{N-1} c_je^{i 2 \pi \frac{j k}{N}} \right) = \frac{1}{N} \sum_{k= 0}^{N - 1}{(\mathscr{F}\beta)_k (\mathscr{F}^{-1} c)_k}\,,
\end{equation}
we see that each Fourier component of the background contributes to the error in proportion to the (inverse) Fourier component of the modulation waveform at that frequency. In the laboratory, the Fourier spectrum of the background is often dominated by the low frequency components, typified by the well-known $1/f$ power spectrum. In general terms, then, it is often desirable to have the Fourier spectrum of the modulation waveform reside at as high a frequency as possible. The simple modulation $c_j = (-1)^j$ achieves this, indeed that is the square-wave version of the standard lock-in detection scheme \cite{Cosens34, Michels41,Michels49}. We shall see, though, that switching waveforms  more complex than this fast square wave have some advantages when we come to consider multi-channel detection in section \ref{Sec:multi}.

Equation \ref{eq:fourier} describes how a particular time-dependent background, $\beta_j$, will produce an error in a particular measurement of $\hat\alpha$. To properly consider the impact of \emph{noisy} backgrounds we need to consider what happens when we repeat the measurement a number of times. A noisy background is one that differs from one experimental realisation to the next, in contrast to a \emph{reproducible} background which is the same for each experimental realisation. We extend our treatment to multiple experimental realisations by allowing $\beta_j$ to be a stochastic variable. To evaluate the effectiveness of a switching waveform it is then necessary to consider the statistical properties of the error, $r_\alpha$, averaged over many experimental realisations.

We begin by limiting ourselves to purely noisy backgrounds, that is backgrounds for which the expected value of $\beta_j$ is  $\mathbb{E}(\beta_j) = 0$.
We consider backgrounds where this limitation does not hold in section \ref{Sec:repeat}. For these purely noisy backgrounds it is easy to see from equation \ref{eq:demoderr} that $\mathbb{E}(r_\alpha) = 0$, which is to say that our estimates of $\alpha$ will be unbiased. They will, though, be noisy as $\beta_j$ changes from one experimental realisation to the next. To quantify this noise let us calculate the standard deviation, $\sigma(r_\alpha)$, which gives the statistical uncertainty in our measurement of $\alpha$. This is given by 
\begin{equation}
\label{eq:sd}
\sigma(r_\alpha)^2 =  \mathbb{E}(r_\alpha r_\alpha^\star)= \mathbb{E} \left(  \frac{1}{N} \sum_{k= 0}^{N - 1}{(\mathscr{F}\beta)_k (\mathscr{F}^{-1} c)_k}  \frac{1}{N} \sum_{l = 0}^{N - 1}{(\mathscr{F}\beta)_l^\star (\mathscr{F}^{-1} c)_l^\star }\right)\,.
\end{equation}
In the first step here, we have used the fact that $r_\alpha$ has zero mean.

It is common to characterise noise by a power spectrum, which defines the (real) amplitude $\tilde{\beta}_k$ of each Fourier component, assumed to be constant for each experimental realisation. The phase $\phi_k$ of each component must vary in some way between realisations for this to be called noise. Let us make the simplifying assumption that the phases for each component are independently drawn from a uniform distribution over $[0,2\pi)$ for each experimental realisation. Although this noise model is quite restricted, it does provide a reasonable approximation to typical laboratory noise and it simplifies equation \ref{eq:sd} in an instructive way. Inserting $(\mathscr{F}\beta)_k = \tilde{\beta}_k e^{i \phi_k}$ into equation\,\ref{eq:sd}, and writing the expected value operator explicitly as a sum, we obtain
\begin{eqnarray*}
\sigma(r_\alpha)^2 =  \lim_{M \to \infty} \frac{1}{M}  \frac{1}{N^2} \sum_{n = 1}^M  \sum_{k = 0}^{N - 1} \sum_{l = 0}^{N - 1}\tilde{\beta}_k e^{i \phi_{k,n}} (\mathscr{F}^{-1} c)_k \tilde{\beta}_l e^{-i \phi_{l,n}} (\mathscr{F}^{-1} c)_l^{\star} \\
 =  \frac{1}{N^2}  \sum_{k = 0}^{N - 1} \sum_{l = 0}^{N - 1} (\mathscr{F}^{-1} c)_k  (\mathscr{F}^{-1} c)_l^{\star}  \tilde{\beta}_k \tilde{\beta}_l \left( \lim_{M \to \infty} \frac{1}{M} \sum_{n = 1}^M e^{i (\phi_{k,n} - \phi_{l,n})} \right)\ ,
\end{eqnarray*}
where $\phi_{j,n}$ is the $n$th realisation of the stochastic variable $\phi_j$. The quantity in parentheses is zero unless $k = l$, hence the sum simplifies to
\begin{equation}
\sigma(r_\alpha) = \frac{1}{N} \sqrt{ \sum_{k= 0}^{N - 1} |(\mathscr{F}^{-1} c)_k|^2\   \tilde{\beta}_k^2 }\ .
\end{equation}
We see that the individual frequency contributions sum in quadrature for this class of noise.

\section{Multichannel modulation and demodulation}
\label{Sec:multi}

In section \ref{Sec:single} we have considered how to find the modulation strength $\alpha$ when one parameter is modulated by control parameter $c$. Now we extend this to the case of multiple modulations. To help establish the notation, let us consider the signal when there are two control parameters, $c^{(1)}$ and $c^{(2)}$,
\begin{equation}
s_j = \beta_j + \alpha^{(1)} c_j^{(1)} + \alpha^{(2)} c_j^{(2)} + \alpha^{(1,2)} c_j^{(1)} c_j^{(2)}\ .
\end{equation}
In addition to the responses $\alpha^{(1)}$ and $\alpha^{(2)}$, there is also the  bilinear response $\alpha^{(1,2)}$ to the product of control parameters. Put another way, this part of the signal is correlated with the \emph{relative} sign of the two control parameters. Similarly, with three modulations, there is a trilinear term as well as the bilinear and linear ones. When $P$ parameters are modulated by switching waveforms $c^{(k)}$, where $k = 1 \ldots P$, the signal generalises to
\begin{equation}
\label{eq:mcsig}
s_j = \beta_j + \sum_{q \subseteq \{1 \ldots P\}\atop  q \neq \emptyset} \alpha^{q} \prod_{k \in q} c_j^{(k)}\ =  \beta_j+ \sum_{q \subseteq \{1 \ldots P\}\atop q \neq \emptyset} \alpha^{q} w_j^{q}\,.
\end{equation}
Here $q$ labels all the subsets of $\{1 \ldots P\}$, i.e., all the linear, bilinear, trilinear, etc. terms. For convenience, the last equality renames the product of control parameters in term $q$ as $w_j^{q}$ ($q$ being a label, not a power), which we sometimes refer to as an \emph{analysis waveform}. 

With several experimental parameters $k$, each modulated by a different waveform $c^{(k)}$ to produce a string of $N$ signals $s_{j}$, the response $\alpha^{(k)}$ to each individual  modulation is estimated by the generalisation of equation\,\ref{eq:extract}:
\begin{equation}\label{eq:mcextract}
\hat\alpha^{(k)} = \frac{1}{N} \sum_{j = 0}^{N - 1}{s_j c_j^{(k)}}\ .
\end{equation}
This relies on the use of switching patterns $c^{(k)}$ that are mutually orthogonal.  As before we will require the switching waveforms to be balanced, so that constant backgrounds are rejected.

The same strategy also applies to the multi-linear responses, allowing the further generalisation  
\begin{equation}\label{eq:mmcextract}
\hat\alpha^{q} = \frac{1}{N} \sum_{j = 0}^{N - 1}{s_j w_j^{q}}\,,
\end{equation}
provided the analysis waveforms $w^{q}$ are also mutually orthogonal. In order that these multi-linear responses also reject constant backgrounds we will require the analysis waveforms to be balanced, which does not automatically follow from the switching waveforms being balanced. We will call a set of switching waveforms \emph{completely balanced} if all products of them (i.e. all analysis waveforms) are balanced.

Harrison, Player and Sandars \cite{Harrison71} have presented a simple scheme for generating a set of mutually orthogonal, completely balanced switching waveforms, and a compact notation for naming them. They start with a basic set of square waves,
\begin{equation}
b(m)_j = (-1)^{\lceil \frac{j+1}{2^m} \rceil + 1}\ ,
\end{equation}
where $\lceil \,\rceil$ indicates the ceiling function. As in section \ref{Sec:single}, the integer $j$ specifies the time step and runs from zero to $N-1$, where $N$ is the total number of steps in the switching sequence, constrained to be an even integer power of 2. The integer label $m$ ranges from zero to $\log_2(N/2)$. For instance, when $N=8$, we have three basic square waves
\begin{eqnarray}
b(0) = \{+1, -1, +1, -1, +1, -1, +1, -1\}&\\
b(1) = \{+1, +1, -1, -1, +1, +1, -1, -1\}&\\
b(2) = \{+1, +1, +1, +1, -1, -1, -1, -1\}&.
\end{eqnarray}
Switching waveforms $c$ are constructed by multiplying any combination of basic square waves together. It is convenient for these to be identified by a binary code that indicates which basic square waves are included, with the $n$th least significant bit indicating whether the $n$th basic waveform is included. 
For example,
\begin{eqnarray}
c(010) =& b(1) &= \{+1, +1, -1, -1, +1, +1, -1, -1\}\\
c(101) =& b(2) \times b(0) &= \{+1, -1, +1, -1, -1, +1, -1, +1\}\\
c(110) =& b(2)\times b(1) &= \{+1, +1, -1, -1, -1, -1, +1, +1\}\ .
\end{eqnarray}
We will find it useful later to introduce an optional, over-all inversion, indicated by an asterisk. For example, $c(101^*) =-c(101)$.

The basic waveforms are all balanced. The symmetry of the HPS waveforms also ensures that multiplying any waveform by a basic waveform flips an equal number of positive and negative entries, leaving the result balanced. In this way, all non-trivial waveforms generated with the HPS technique are balanced \footnote{The HPS waveforms constitute only a fraction of all the balanced waves of a given length. This fraction becomes vanishingly small for long patterns but nevertheless there are enough for practical purposes. For a pattern of (even) length $N$, the number of non-trivial HPS waveforms is $N-1$. By contrast, the number of balanced waveforms is $N! / (\frac{N}{2}!)^2$, that being the number of distinct ways to assign the value $+1$ to $N/2$ of the sites (and $-1$ to the rest).}. Note that the trivial pattern of constant $+1$ is a valid HPS waveform, which we label $c(0)$, and while it is not balanced, it is useful to include in the discussion as it can be used to extract the average value of the background.

In order to extract all the responses from a signal, all analysis waveforms $w^{q}$ must be mutually orthogonal. The construction of the HPS waveforms ensures that all distinct waveforms are orthogonal, meaning it is sufficient to ensure that all analysis waveforms are distinct. Note well that this is a stronger constraint than ensuring that all \emph{switching} waveforms are distinct. This point was not addressed by HPS \cite{Harrison71}, but is of great practical relevance when it comes to choosing the set of modulation waveforms. For example, let us choose three modulation waveforms $c^{(1)}=c(001)$, $c^{(2)}=c(011)$ and $c^{(3)}=c(010)$. These are mutually orthogonal, but the analysis waveform $w^{(1,2)} = c^{(1)}c^{(2)}$ is the same as $c^{(3)}$ and therefore we cannot separate the bilinear response  $\alpha^{(1,2)}$ from the linear response $\alpha^{(3)}$ with this choice of modulation waveforms.

Requiring all the analysis waveforms to be distinct imposes a minimum length on the waveform, which depends on the number of parameters being modulated. Given $P$ parameters, each modulated with a different switching pattern $c^{(k)}$, the number of independent analysis waveforms $w^{q}$ is $2^{P}$ \footnote{There are $P$ linear responses, $P(P-1)/2!$ bilinear responses, $P(P-1)(P-2)/3!$ trilinear, etc., terminating with one response to the product of all $p$ modulations, plus the trivial waveform $c(0)$. Summing the number of responses we obtain a total of $2^P$. }. Since the signal is made up of $N$ numbers, $s_j$, the linear combinations of its elements can only represent $N$ independent variables, and therefore a full analysis is only possible if the length $N$ of the switching pattern is at least  $2^P$. This means that the number of digits in the longest binary HPS code must  equal or exceed the number of parameters being modulated.
 
To test whether all analysis waveforms resulting from a given set of switching waveforms are distinct and orthogonal, one could simply generate them and check. In practice, however, we normally test the waveforms in a different way that is equivalent but easier to implement. With $P$ switched parameters, the set of switches can be in any one of $2^P$ combined states in each time interval. If the set of switching waveforms is such that each of these $2^P$ states is visited an equal number of times in the $N$ measurements, then all of the analysis waveforms are distinct and orthogonal. In the example above of $c^{(1)}=c(001)$, $c^{(2)}=c(011)$ and $c^{(3)}=c(010)$, only 4 of the 8 combined switch states are accessed, specifically $(c^{(1)}, c^{(2)}, c^{(3)}) = \{(-1,-1,+1), (-1,+1,-1),(+1,-1,-1),(+1,+1,+1)\}$, indicating that these switching waveforms do not result in a full set of orthogonal analysis waveforms.

We will employ HPS waveforms exclusively in the rest of this paper.

\section{Reproducible backgrounds}
\label{Sec:repeat}

In the latter part of section\,\ref{Sec:single} we considered a background $\beta$ that varies purely randomly, that is without any consistent, repeatable part. We showed that although the measurement is perturbed by this background, $\hat\alpha$ converges to the true value $\alpha$ in the long-term average. In this section we allow some part of the background to vary in the same way over each set of $N$ measurements. We will see that this gives a biased result when using the detection method described so far, and we show how a modification of the method can remove this bias.

To give a concrete example of such a background we consider a measurement recently made in our laboratory \cite{Hudson11,Kara12}. In this experiment a molecular beam is subjected to a number of manipulations by static electric and magnetic fields, rf magnetic fields, and laser light, after which we measure the number of molecules in a particular quantum state.  We modulate these fields in a number of ways, and in particular, we switch the signs of the static electric and magnetic fields. We seek to detect a modulation strength associated with the relative sign of the electric and magnetic fields, which would indicate that the charge distribution within electrons is not spherical. We modulate 9 experimental parameters using the scheme described in this paper. Our waveforms have length $N=4096$, and making a single \emph{block} of $4096$ measurements takes a few minutes. We then repeat this experiment, for months at a time, amassing many thousands of blocks in order to reduce the final measurement uncertainty. Noise sources include shot noise and electronic noise in the detection system, fluctuations of the beam intensity, and instability of the various fields. Using the techniques described above we reduce the impact of that noise through careful choice of the modulation waveforms. However, part of the varying background is reproducible. For example, the molecular source heats up during each block of measurements, resulting in a drift of the signal $s_j$ as the experiment progresses. There is a pause between each block where various parameters of the equipment are optimised and during this pause the source cools down again. The consequent repeating background introduces a bias that would invalidate the measurement, but we are able to suppress the bias by the method we now describe.

Let us consider a background which is the sum of a reproducible part $\gamma$ and a purely random part. We define the bias $\epsilon^{q}$ as the error in measuring the parameter $\alpha^{q}$, averaged over many experimental realisations. Equation \ref{eq:mmcextract} and equation \ref{eq:mcsig} then give
\begin{equation}
\label{eq:bias}
\epsilon ^{q}= \mathbb{E}(\hat\alpha^{q} - \alpha^{q}) = \mathbb{E}\left[\frac{1}{N} \sum_{j = 0}^{N - 1}{\gamma_j w^{q}_j}\right] = \frac{1}{N} \sum_{j = 0}^{N - 1}{\gamma_j w^{q}_j}\ ,
\end{equation}
where we have used the linearity of equation \ref{eq:mmcextract} and the fact that the random part of the background generates no bias. The last equality is intended to show the problem clearly: in each experimental realisation the error due to $\gamma$ is the same, producing a bias that does not average away.

To get out of this pickle we can vary the switching waveforms between experimental realisations. In this case the last equality in equation \ref{eq:bias} does not hold, and we must write instead
\begin{equation}
\epsilon ^{q}= \frac{1}{N} \sum_{j= 0}^{N - 1}{\gamma_j\mathbb{E} \left[ w^{q}_j\right] }\,.
\end{equation}
We can see that this offers a solution to the problem: the bias will vanish if the waveform, averaged over all experimental realisations, is orthogonal to the repeating part of the background. The simplest way to arrange this for all $\gamma$ is to randomise the waveforms, making $\mathbb{E} \left[ w^{q}_j \right]=0$.  Randomising the waveforms makes sure that the error from the reproducible background is different in each measurement, in effect converting the bias into noise. There is a pleasing symmetry to this: by introducing structured waveforms we were able to reduce the impact of noisy backgrounds on our measurement, and  we see that by introducing randomness into the waveforms we are able to reduce the impact of structured backgrounds.

\section{Partially randomised waveforms}
\label{Sec:random}

\begin{figure}
\center
\includegraphics[width=10cm]{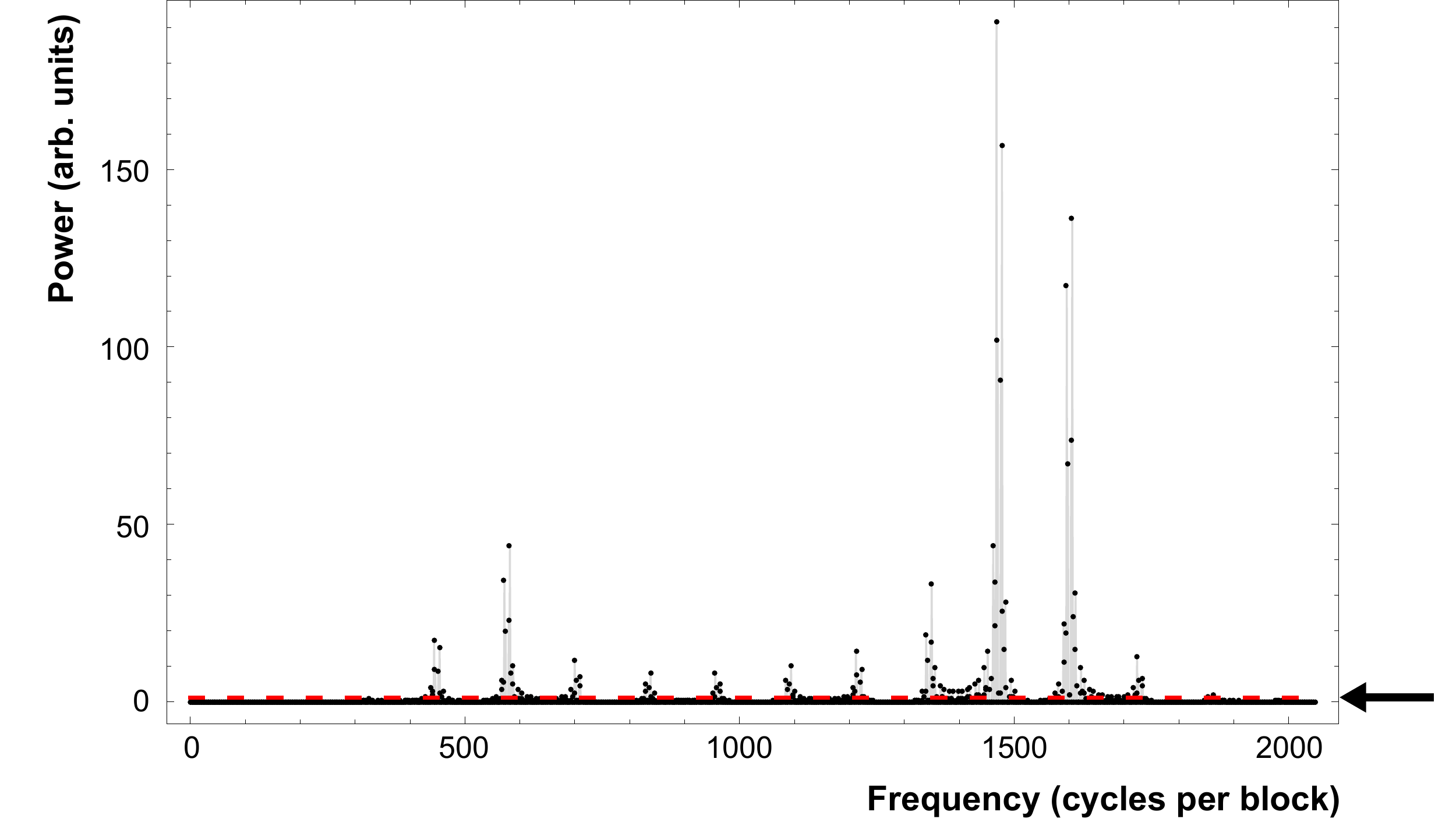}
\caption{The power spectrum of a single switching waveform, $c(101100110111)$. The dashed line and arrow indicate the power level that would result from spreading the power equally over all Fourier components.}
\label{fig:single}
\end{figure}

Our aim is to randomise the waveforms partially: enough to make sure that measurement bias from reproducible backgrounds is suppressed, while retaining the noise rejecting properties of the lock-in scheme. We therefore generate new waveforms for each experimental realisation by randomly generating HPS codes, but with constraints to ensure that they have the frequency properties we desire. The HPS waveform notation provides a particularly simple way to do this. 

Let us consider again the experiment described in section \ref{Sec:repeat}, which uses switching waveforms of length $N=4096$, specified by 12-bit HPS codes. To select suitable waveforms for this experiment we first generate a set of 12-bit codes, one for each switch, choosing the bits at random \footnote{In fact, we do not randomly generate all of the codes. Two of the switch channels in our experiment must be switched with particular sequences due to hardware constraints. The procedure we describe for generating partially randomised waveforms is used for generating the other seven switching waveforms.}. We then check each analysis channel of particular interest to see whether it satisfies two constraints. First, its HPS code must not have more than one zero in the last four bits. This ensures that the corresponding analysis waveform will reject low frequency noise as a result of its strong high-frequency content. Second, there must be no more than four zeros in the leading eight bits representing the slower basic waveforms. This ensures that the analysis waveform has a complex structure, needed to suppress the residual error due to slow variations in the background, as discussed by HPS \cite{Harrison71}. If either test is failed, we generate a new set of codes and repeat until the tests are passed. Finally, we chose at random whether to invert each switching waveform, with probability 1/2.

\begin{figure}
\center
\includegraphics[width=10cm]{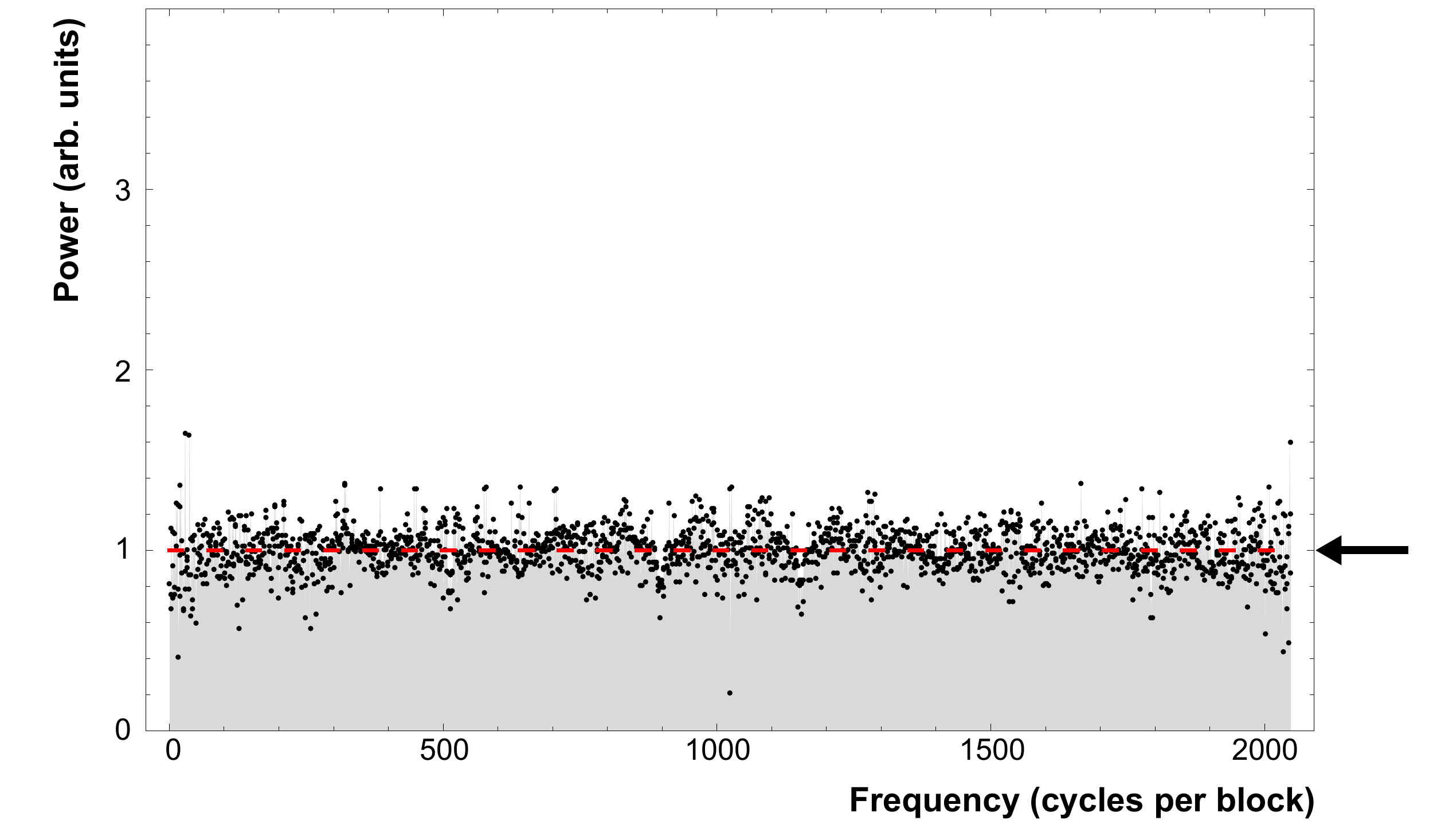}
\caption{The average power spectrum of completely randomised HPS waveforms, with random inversion. Again, the dashed line and arrow show the power level that would result from spreading the power over all Fourier components --- the white noise level --- which we see approximates the resulting power spectrum well. The plot was generated by averaging 10,000 waveforms generated with randomly chosen HPS codes.}
\label{fig:random}
\end{figure}

This partial randomisation scheme completely eliminates bias in the measurement. At any step in the sequence of $N$ time bins, every analysis waveform is randomised, with equal probability of being -1 or +1 because of the randomly chosen inversions. This satisfies the requirement that $\mathbb{E} \left[ w^{q}_j \right]=0$. By randomising the waveform codes between each experimental realisation we also ensure that no one analysis waveform performs consistently poorly, as would be likely to happen if fixed codes were used.

To understand the impact of the partial randomisation scheme, it is instructive to look at the average power spectra of some waveforms. Figure \ref{fig:single} shows the power spectrum of a particular 12-bit HPS waveform, $c(101100110111)$. We see that this has most of its power concentrated at high frequency, so it is effective in rejecting low frequency noise. However, as we have seen, we will be susceptible to bias if we use this waveform for every measurement. At the opposite extreme, figure \ref{fig:random} shows the average power spectrum for $10^4$ HPS waveforms chosen randomly without any constraint on the HPS code. While these waveforms will yield an unbiased result, they do not provide good noise rejection as the average power spectrum is completely white.

\begin{figure}
\center
\includegraphics[width=10cm]{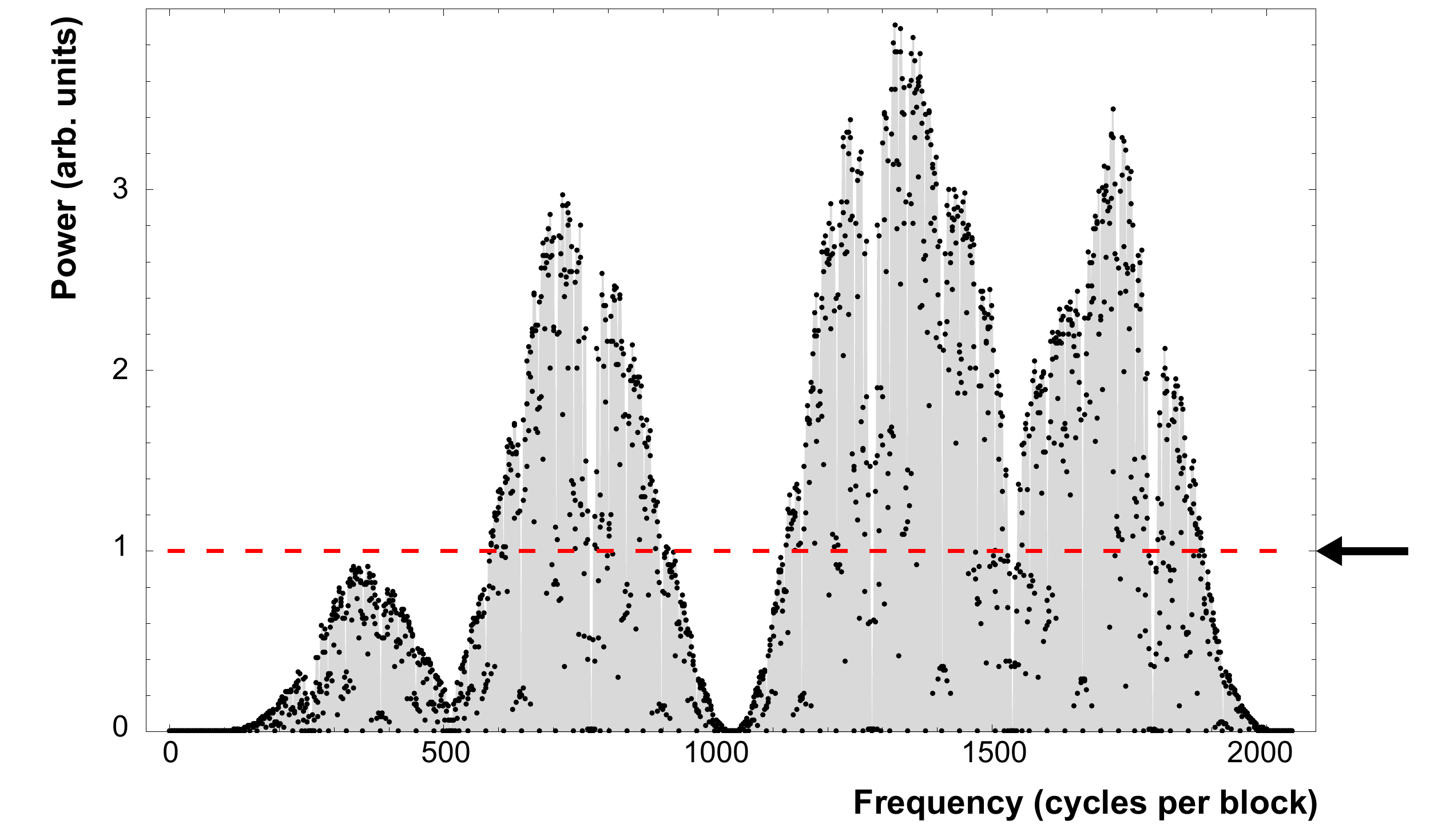}
\caption{Average power spectrum of waveforms selected by the partial-randomisation algorithm described in the text. The dashed line and arrow show the white noise spectrum. The plot was generated by averaging the power spectra of 10,000 waveforms generated using the algorithm in the text.}
\label{fig:partial}
\end{figure}

Figure \ref{fig:partial} shows the average power spectrum of a waveform generated using the algorithm described above. We see that this waveform offers the best of both worlds. The average power is spread over a broad range of the higher frequencies while low freqency components are almost completely absent. Thinking in the time-domain, if we were to string together all the partially randomised waveforms chosen for a given analysis channel, the switching pattern would be high-frequency periodic in the short term but random and uncorrelated in the long term. It is this combination of short-term periodicity to reject noise, together with long term randomness to reject repetitive backgrounds, that makes the scheme effective.

\section{Summary and conclusions}
\label{Sec:conclusions}
We have shown that the multi-channel lock-in detection scheme introduced by HPS yields information on all the responses, but only after minimum-length and completeness requirements are applied to the choice of modulation waveforms. A method is given for testing the suitability of the waveform set. We have shown how to evaluate the standard deviation of the measurement results in the presence of random noise. Finally, we have pointed out that many experiments also have a reproducing pattern of background variation that can bias the results of the measurement. We have described a scheme for partially randomising the modulation waveforms in such a way that this bias averages to zero and merely contributes to the noise.

\ack
This work was supported by the Royal Society, the UK EPSRC and STFC and by the European Commission.

\end{document}